\documentclass[preprint2,flushrt]{aastex}

\received{2001 March 6}
\accepted{2001 May 17}

\def\mbar{\ifmmode\overline{m}\else$\overline{m}$\fi}
\def\Mbar{\ifmmode\overline{M}\else$\overline{M}$\fi}
\def\mibar{\ifmmode\overline{m}_I\else$\overline{m}_I$\fi}
\def\MIbar{\ifmmode\overline{M}_I\else$\overline{M}_I$\fi}
\def\Nbar{\ifmmode\overline{N}\else$\overline{N}$\fi}
\def\hst{\emph{HST}}
\def\ho{\ifmmode H_0\else$H_0$\fi}
\def\sna{SNIa}

\def\dmod{\ifmmode(m{-}M)_0\else$(m{-}M)_0$\fi}
\def\mM{\ifmmode(m{-}M)_0\else$(m{-}M)_0$\fi}
\def\vi{\ifmmode(V{-}I)\else$(V{-}I)$\fi}
\def\viz{\ifmmode(V{-}I)_0\else$(V{-}I)_0$\fi}
\def\EBV{\ifmmode E_{B-V}\else$E_{B-V}$\fi}
\def\kmsMpc{\hbox{$\,$km$\,$s$^{-1}\,$Mpc$^{-1}$}}
\def\kms{\hbox{$\,$km$\,$s$^{-1}$}}

\shorttitle{SBF Distances to \sna}
\shortauthors{Ajhar et al.}

\begin{document}

\title{Reconciliation of the Surface Brightness Fluctuations \\
and Type~Ia Supernovae Distance Scales\altaffilmark{1}}

\author{Edward A. Ajhar\altaffilmark{2}, John L. Tonry\altaffilmark{3}, 
John P. Blakeslee\altaffilmark{4}, Adam G. Riess\altaffilmark{5}, \\
and Brian P. Schmidt\altaffilmark{6}}

\altaffiltext{1}{Based on observations with the NASA/ESA \emph{Hubble
Space Telescope,} obtained at the Space Telescope Science Institute,
which is operated by the Association of Universities for Research in
Astronomy (AURA), Inc., under NASA contract NAS5-26555. These 
observations are associated with proposal IDs~8212, 5990, and~6587.}
\altaffiltext{2}{Kitt Peak National Observatory, National Optical Astronomy
Observatories, P.~O. Box 26732, Tucson, AZ 85726;
{ajhar@physics.miami.edu}.  Current address: Department of Physics,
University of Miami, P.~O.\ Box 248046, Coral Gables, FL 33124-8046} 
\altaffiltext{3}{Institute for Astronomy, University of Hawaii,
2680 Woodlawn Drive, Honolulu, HI 96822; {jt@ifa.hawaii.edu}}
\altaffiltext{4}{Department of Physics and Astronomy,
Johns Hopkins University, 3701 San Martin Drive, Baltimore, MD 21218; {jpb@pha.jhu.edu}}
\altaffiltext{5}{Space Telescope Science Institute, 3700 San Martin Drive,
Baltimore, MD 21218; {ariess@stsci.edu}}
\altaffiltext{6}{Mount Stromlo and Siding Spring Observatories,
Private Bag, Weston Creek PO 2611, Australia; {brian@mso.anu.edu.au}}

\begin{abstract}

We present \emph{Hubble Space Telescope} measurements of surface
brightness fluctuations (SBF) distances to early-type galaxies that
have hosted Type~Ia supernovae (\sna).  The agreement in the relative
SBF and \sna\ multicolor light curve shape (MLCS) and delta-$m_{15}$
(dm15) distances is excellent.  There is no systematic scale error with
distance, and previous work has shown that SBF and \sna\ give 
consistent ties to the Hubble flow.  
However, we confirm a systematic offset of $\sim$ 0.25~mag
in the distance zero points of the two methods, and we trace this offset
to their respective Cepheid calibrations.  SBF has
in the past been calibrated with Cepheid distances from the \ho\
Key Project team, while \sna\ have been calibrated with Cepheid
distances from the team composed of Sandage, Saha, and collaborators.
When the two methods are calibrated in a consistent way,
their distances are in superb agreement.  Until the conflict over the ``long''
and ``short'' extragalactic Cepheid distances among many galaxies is
resolved, we cannot
definitively constrain the Hubble constant to better than $\sim10$\%,
even leaving aside the additional uncertainty in the distance to the
Large Magellanic Cloud, common to both Cepheid scales. 
However, recent theoretical
SBF predictions from stellar population models favor the Key Project
Cepheid scale, while the theoretical \sna\ calibration lies between
the long and short scales.
In addition, while the current SBF distance to M31/M32
is in good agreement with the RR~Lyrae and red giant branch distances,
calibrating SBF with the longer Cepheid scale would introduce 
a 0.3~mag offset with respect to the RR~Lyrae scale.
\end{abstract}

\keywords{galaxies: distances and redshifts ---
cosmology: distance scale --- supernovae: general}

\section{Introduction}

The single most important parameter in an expanding universe model is the
rate of expansion, that is, the Hubble constant~\ho.  
From the ages of the oldest stars, to the masses of galaxies and clusters,
to the cosmic baryon density predicted by big-bang nucleosynthesis,
\ho\ underlies everything.  Knowledge of \ho\ is essential for
relating physical scales within our Galaxy to extragalactic phenomena:
it sets the age, distance, and mass scales for the universe.

The most straightforward way of determining \ho\ is by measuring
calibrated galaxy distances well out into the Hubble flow, where
peculiar velocities become unimportant.  Two of the highest precision
methods for measuring extragalactic distances are Type~Ia supernovae
(\sna) and surface brightness fluctuations (SBF).  Both methods
typically achieve distance accuracies of 10\% or better.  While \sna\
have a long history as standard candles, only recently have the
empirical calibration of their luminosities in terms of the decline rate
(parametrized by ``dm15,'' the difference between the magnitudes
at peak and fifteen days after peak)
\citep{phillips.1993} and the use of multicolor light curve shapes 
(MLCS) to constrain dust extinction
\citep{riess.etal.1998} made them
outstandingly accurate.  Similarly, the SBF method showed great
promise from its inception
\citep*{tonry.schneider.1988,tonry.ajhar.luppino.1990}, but only
recently has it been well-calibrated in terms of stellar population
(\citealt{sbf.i}, \citeyear{sbf.ii}, hereafter SBF-I and SBF-II;
\citealt{ajhar.etal.1997}) \defcitealias{sbf.i}{SBF-I}
\defcitealias{sbf.ii}{SBF-II} and extensively modeled from a
theoretical perspective
\citep*{worthey.1993a,worthey.1993b,liu.charlot.graham.2000,blakeslee.vazdekis.ajhar.2001}.

Curiously, SBF and \sna\ have often been associated with opposing
camps in the traditional distance scale controversy.  The \sna\ method
has seemed to favor the ``long'' distance scale with $\ho\approx60$
\kmsMpc\
\citep{riess.etal.1998,hamuy.etal.1996,sandage.etal.1996,saha.etal.1999},
while SBF has seemed to favor the ``short'' distance scale with
$\ho\approx80$ \kmsMpc\ (\citetalias{sbf.i};
\citealt{lauer.etal.1998}).  Several years ago, it was possible to
discount \ho\ from SBF because ground-based SBF distances only reached
to $\sim3000$ \kms, requiring a tie to the Hubble flow via some other
distance estimator, but this extra step has not been a part of the
more recent SBF \ho\ determinations.

Over the past few years there has been considerable convergence on
\ho.  For instance, \citet[hereafter
SBF-III]{sbf.iii}\defcitealias{sbf.iii}{SBF-III} find $\ho=74\pm4\pm7$
by correcting the peculiar velocities of the SBF survey galaxies
\citep[hereafter SBF-IV]{sbf.iv}\defcitealias{sbf.iv}{SBF-IV}
according to the gravity field derived from the \textit{IRAS} 1.2\,Jy
redshift survey \citep{fisher.etal.1995}, and \citet{jensen.etal.2001}
find $\ho=72\pm3\pm7$ \kmsMpc\ from $H$-band \textit{Hubble Space
Telescope} (\hst) NICMOS SBF distances to 6 galaxies beyond 7000~\kms.
Using the \sna\ method, \citet{gibson.stetson.2001} find
$\ho=73\pm2\pm7$ \kmsMpc\ based on nine \sna\ hosts with Cepheid
distances. The \ho\ Key Project group (hereafter KP) has reported
final values of $\ho=70\pm5\pm6$ from SBF and $\ho=71\pm2\pm6$ from
\sna\ \citep{freedman.etal.2001}. Moreover, the results of
\citetalias{sbf.iii} and \citet{riess.etal.1997} show that SBF and
\sna\ trace the same local velocity field, and therefore imply
a consistent tie to the Hubble flow. 

However, the Sandage-Saha group (hereafter S\&S) find a significantly
smaller $\ho=58.5\pm6.3$ \kmsMpc\ \citep{parodi.etal.2000}, based on what
are essentially the same \sna\ and Cepheid data.  The disagreement over
the value of \ho\ from \sna\ is due to a number of subtle differences in
the data treatment, including the Cepheid measurements themselves, as
discussed by \citet{parodi.etal.2000} and \citet{gibson.etal.2000}.

Faced with this controversy over \ho\ derived from SBF and \sna, we 
undertook to obtain new SBF measurements in galaxies which have hosted \sna.
Because the early-type galaxies best suited for SBF are rare
in comparison to late-tape galaxies and because well-measured \sna\
are much rarer still, we were forced to distances beyond 2000\kms,
which requires \hst\ for high quality SBF distances.
In this paper we present new SBF measurements with \hst/WFPC2 for
seven galaxies which have hosted \sna, and we also report ground-based
SBF distances for seven other \sna\ hosts.  
These data allow us to evaluate the agreement between SBF and \sna\
relative distances over an unprecedented distance range and with
enough galaxies to provide solid statistics.

\section{\hst\ Observations}

The primary data set comprises \textit{Hubble Space Telescope} images
of five galaxies observed from 1999 July through 2000 June with the
Wide Field and Planetary Camera~2 (WFPC2)
\citep{holtzman.etal.1995a,holtzman.etal.1995b} through our GO program
8212.  In addition we report measurements of NGC~1316 taken from
program 5990 and NGC~5061 taken from program 6587.  We selected
elliptical and lenticular galaxies hosting normal \sna\ for which 
suitable \sna\ data existed and which were near enough to allow
good SBF measurements in relatively few \hst\ orbits.  Our candidate
selection has allowed us to extend the distance modulus range of
overlap between these two methods by 2.5~mag.  The two methods now
have a common distance range of 10--73~Mpc.

Table~\ref{tblgxys} lists each galaxy in column 1 with its
corresponding supernova in column 2.  Column 3 lists the velocity 
(km~s$^{-1}$) transformed to the CMB frame, while the WFPC2 exposure
times (in seconds) are listed in column 4 and $B$-band Galactic
extinctions from \citet*[hereafter SFD]{sfd}\defcitealias{sfd}{SFD}
are in column 5.  The exposure times were chosen for each galaxy to
obtain sufficient signal-to-noise for good SBF measurements.  The
typical exposure sequence for a given galaxy consisted of several
dithered exposures in the F814W ($I$) filter to provide cosmic ray
rejection and better flattening.  The galaxies were always centered in
the Planetary Camera (PC) CCD.  The SBF analysis was conducted on the
PC images alone, while the sky levels and total galaxy magnitudes were
estimated by constructing a mosaic of the PC and the three WF
images---forming the usual chevron.

\begin{deluxetable}{llcrcccc}
\tabletypesize{\scriptsize}
\tablewidth{0pt}   
\tablecaption{Galaxy Properties and SBF Measurements}
\tablehead{
\colhead{Galaxy} &
\colhead{\sna} &
\colhead{$v_\mathrm{CMB}$} &
\colhead{$t_\mathrm{exp}$} &
\colhead{$A_B$} &
\colhead{\viz} &
\colhead{\mibar} &
\colhead{\Nbar} \\
\colhead{(1)} &
\colhead{(2)} &
\colhead{(3)} &
\colhead{(4)} &
\colhead{(5)} &
\colhead{(6)} &
\colhead{(7)} &
\colhead{(8)} }
\startdata%
\multicolumn{8}{c}{WFPC2 F814W Data} \\ \tableline
E352-057&1992bo&5331 &23,800& 0.12 & $1.153\pm0.015$ & $32.61\pm 0.10$ & 20.28 \\
N1316 & 1980N & 1657 & 1860 & 0.09 & $1.170\pm0.018$ & $29.60\pm 0.19$ & 22.66 \\
N2258&1997E&3961&16,200&1.13\tablenotemark{a}&$1.22\pm0.03$&$32.15\pm 0.10$&22.38\\
N2962 & 1995D & 2450 & 6900 & 0.25 & $1.176\pm0.017$ & $31.10\pm 0.09$ & 21.06 \\
N5061 & 1996X & 2353 & 1800 & 0.30 & $1.099\pm0.023$ & $30.34\pm 0.15$ & 21.65 \\
N5308 & 1996bk& 2147 & 4000 & 0.08 & $1.175\pm0.030$ & $30.97\pm 0.09$ & 20.73 \\
N6495 & 1998bp& 3144 & 9200 & 0.34 & $1.219\pm0.015$ & $31.86\pm 0.10$ & 20.83 \\
\tableline
\multicolumn{8}{c}{Ground-based SBF Survey Data} \\ \tableline
N0524 & 2000cx& 2091 &\nodata& 0.36 & $1.221\pm0.010$ & $30.48\pm 0.19$ & 21.59 \\
N1316 & 1980N & 1657 &\nodata& 0.09 & $1.132\pm0.016$ & $29.83\pm 0.15$ & 22.66 \\
N1380 & 1992A & 1732 &\nodata& 0.08 & $1.197\pm0.019$ & $29.70\pm 0.15$ & 20.77 \\
N3368 & 1998bu& 1252 &\nodata& 0.11 & $1.145\pm0.015$ & $28.32\pm 0.20$ & 20.22 \\
N4374 & 1991bg& 1375 &\nodata& 0.17 & $1.191\pm0.008$ & $29.77\pm 0.09$ & 21.97 \\
N4526 & 1994D &  949 &\nodata& 0.10 & $1.188\pm0.021$ & $29.57\pm 0.17$ & 21.13 \\
N5128 & 1986G &  812 &\nodata& 0.50 & $1.078\pm0.016$ & $26.05\pm 0.11$ & 21.81 \\
\enddata
\tablenotetext{a}{The \citetalias{sfd} extinction is 0.55, but the one adopted here is based on \Nbar.  See text.}
\label{tblgxys}
\end{deluxetable}

The SBF analysis was essentially routine, following descriptions
discussed elsewhere (\citealp{ajhar.etal.1997},
\citetalias{sbf.i}).  Briefly, for each exposure of each galaxy, we
registered the standard pipeline-reduced CCD images, and then added
them together while statistically eliminating cosmic rays.  The total
summed image for each galaxy was used in the analysis.  We fit a
smooth galaxy model; subtract it from the image; identify, characterize,
and mask all background and foreground objects in the image; and measure 
the power spectrum of the galaxy surface brightness.  We then scale
the galaxy power spectrum to the power spectrum of a suitable point
spread function (PSF), and after allowing for a contribution from
undetected objects in the image, we obtain a value for the
apparent fluctuation magnitude \mbar.

For our measurements we used three composite PSFs taken from images of
$\omega$~Cen in the standard \hst\ PSF program.  We registered (by
whole pixels) and added six to fourteen stars in images taken on three
different dates: 1999/6/6, 1999/8/3, and 2000/3/10.  We measured
\mbar\ using each of these PSFs for each galaxy and took the mean for
our final \mbar.  The agreement among the PSFs was always excellent,
0.08~mag or better. 

Using the chevron images from the \hst\ observations, we used the
program of B.~Barris which fits a Sersic model to the curve of growth
and produces a total magnitude $m_T$ for the galaxy.  Combining this
with our value for \mbar\ we obtain $\Nbar = \mbar -m_T$.  This
extinction independent measure of the absolute luminosity of a galaxy
introduced in \citetalias{sbf.iv} correlates well with a galaxy's
color and \Mbar, and hence can be used with the observed color of a
galaxy to estimate the extinction along its line of sight.

\section{Ground-Based Data}

Obtaining an SBF distance measurement of a galaxy requires knowledge
of its underlying stellar population.  A galaxy's \vi\ color has been
the preferred method for estimating its absolute fluctuation magnitude
\MIbar.  We obtained colors for our program galaxies from the SBF 
survey data and from images taken on photometric nights at the 
UH 2.2-m telescope during 2000 July and
December by B.~Barris and P.~Capak.  These data were reduced in the
usual way, following \citetalias{sbf.i}, and tied to the existing
photometry from the SBF survey.  None of the program galaxies showed a
significant color gradient, and we maintained a good overlap with the
regions covered by the PC.  We found agreement to better than 0.02 mag
between this ground-based photometry and the WFPC2 F814W photometry as
well.  These colors are given in column 6, and \mibar\ and \Nbar\ are
listed in columns~7 and~8, respectively, of Table~\ref{tblgxys}, while
the final SBF distance moduli are presented in \S~\ref{sec:sbfvssna}.

There are a handful of galaxies in the SBF Survey which have hosted
supernovae, and we reproduce the data for these galaxies from
\citetalias{sbf.iv} in Table \ref{tblgxys}.  Two supernovae, SN1986G
and SN1991bg, were subluminous \sna, and although they do not
contribute to the SBF-\sna\ comparison, we include them here for
completeness and for future studies of \sna\ luminosities.

\section{SBF Distances}
\label{secsbfdist}

It is a curious fact that the very small set of early-type galaxies
between 2000 and 5000\kms\ which have hosted \sna\ have a rather large
extinction on average---the median \citetalias{sfd} \EBV\ for our
sample is 0.065 which corresponds to $A_B \sim 0.27$~mag.  We
therefore undertook to calculate the quantity $\Nbar$ discussed in
\citetalias{sbf.iv}, which is an extinction independent measure of a
galaxy's color (accurate to about 0.03~mag for $\vi$) and absolute
luminosity.  The agreement between the values of $\vi$ that we
obtained from ground-based photometry and \citetalias{sfd} extinctions
and the values of $\vi$ that we infer from $\Nbar$ were in superb
agreement in four cases. However, for NGC~2258 we found gross
disagreement between the \citetalias{sfd} corrected ground-based color
of $\vi=1.39$ and the $\Nbar$-estimated color of $\vi=1.22$, and for
NGC~5061 we found a three-sigma disagreement between the
\citetalias{sfd} corrected ground-based color of $\vi=1.10$ and the
$\Nbar$ color of $\vi=1.19$.

NGC~2258 has an estimated $A_B=0.55$~mag from \citetalias{sfd} but is
in a region of highly variable extinction in the \citetalias{sfd}
maps; the $\Nbar$-estimated color suggests that the actual extinction
along the line of sight to the galaxy is $A_B=1.13$~mag.  
The \sna\ analyses described below also call for a larger extinction than
\citetalias{sfd} (but some could be intrinsic to NGC~2258, of course):
the MLCS fit wants $A_B=0.72$~mag and the dm15 fit wants $A_B=0.91$~mag.
As the IRAS pixels were 6\arcmin, it is certainly possible that NGC~2258
is sitting on a local spike of extinction.
In fact, \citet{arce.goodman.1999} have found that the \citetalias{sfd}
map does tend to underestimate
the Galactic extinction in regions with steep extinction gradients.
Because $\vi=1.39$ is far redder than any elliptical galaxy that we
have encountered, we have chosen to adopt the $\Nbar$-estimated
reddening, color, and SBF distance.

NGC~5061 is a less clear-cut case, both because the difference between
\citetalias{sfd} and $\Nbar$ is only three sigma, and because both the
\citetalias{sfd} and $\Nbar$ colors are plausibly correct for this
galaxy.  The velocity dispersion and Mg$_2$ index are consistent with
both colors.  The \citetalias{sfd} extinction for NGC~5061 is $A_B =
0.30$~mag, and the $\Nbar$ extinction is $A_B = -0.02$~mag.  Both our
MLCS and dm15 analyses of SN1996X and that of \citet{salvo.etal.2001}
find $A_B \sim 0.30$~mag to the supernova, and \citet{salvo.etal.2001}
find interstellar Na$\,D$ at the galaxy redshift which would be
appropriate for $A_B \sim 0.30$~mag, but as always, it is possible
that the extinction is local to the supernova.
We have chosen to use the \citetalias{sfd} extinction in this case.
However, we note that if we were to use the $\Nbar$ color, our SBF
distance modulus would drop by 0.33~mag, into better
agreement with the \sna\ distance (after shifting the two sets
of distances to a consistent zero point).

Table~\ref{tbldmodorig} lists the SBF distance moduli for the
program galaxies on the SBF-II zero point, based on the Cepheid
distances tabulated by \citet[hereafter
F00]{ferrarese.etal.2000}\defcitealias{ferrarese.etal.2000}{F00}.
This allows for easy comparison between these SBF distances 
and those from the SBF survey, tabulated in \citetalias{sbf.iv}.
However, note that the \sna\ distances also appearing in
Table~\ref{tbldmodorig} are not tied to the F00 Cepheid distances,
and therefore cannot yet be compared directly with the SBF distances.
The following section discusses the \sna\ distances in detail,
and then \S\ref{sec:sbfvssna} directly compares the SBF and
\sna\ distances after shifting their respective zero points
to a common calibration.

\begin{deluxetable}{llccc}
\tabletypesize{\scriptsize}
\tablewidth{0pt}   
\tablecaption{Distance Moduli with Original Calibrations}
\tablehead{
\colhead{Galaxy} &
\colhead{\sna} &
\colhead{$(m{-}M)_{\mathrm{SBF}}$} &
\colhead{$(m{-}M)_{\mathrm{MLCS}}$} &
\colhead{$(m{-}M)_{\mathrm{dm15}}$} }
\startdata%
\multicolumn{5}{c}{WFPC2 F814W Data} \\ \tableline
E352-057&1992bo& $34.33\pm0.15$ & $34.64\pm0.15$ & $34.83\pm0.11$\\
N1316 & 1980N &  $31.21\pm0.23$ & $31.57\pm0.15$ & $31.77\pm0.10$\\
N2258 & 1997E &  $33.56\pm0.15$ & $34.06\pm0.18$ & $33.91\pm0.15$\\
N2962 & 1995D &  $32.66\pm0.15$ & $32.95\pm0.15$ & $32.95\pm0.15$\\
N5061 & 1996X &  $32.32\pm0.19$ & $32.25\pm0.15$ & $32.39\pm0.10$\\
N5308 & 1996bk&  $32.55\pm0.21$ & $32.50\pm0.21$ & $32.33\pm0.20$\\
N6495 & 1998bp&  $33.16\pm0.15$ & $33.34\pm0.20$ & $33.34\pm0.15$\\
\tableline
\multicolumn{5}{c}{Ground-based SBF Survey Data} \\ \tableline
N0524 & 2000cx& $31.90\pm0.20$ & $32.30\pm0.15$\tablenotemark{a} & $33.13\pm0.14$\tablenotemark{a}\\
N1316 & 1980N & $31.66\pm0.17$ & $31.57\pm0.15$ & $31.77\pm0.10$\\
N1380 & 1992A & $31.23\pm0.18$ & $31.47\pm0.15$ & $31.79\pm0.10$\\
N3368 & 1998bu& $30.08\pm0.22$ & $30.30\pm0.18$ & $30.16\pm0.15$\\
N4374 & 1991bg& $31.32\pm0.11$ &    \nodata     &    \nodata    \\
N4526 & 1994D & $31.14\pm0.20$ & $30.82\pm0.15$ & $31.31\pm0.13$\\
N5128 & 1986G & $28.12\pm0.14$ &    \nodata     & $27.49\pm0.20$\\
\enddata
\tablenotetext{a}{This supernova is a very poor fit (W.~Li et al., in
preparation), and these distances are untrustworthy.  See text.}
\label{tbldmodorig}
\end{deluxetable}

\section{Supernova Distances}

There are a variety of methods for measuring distances to \sna, most
of which use the shape of the SN light curve for estimating the
intrinsic brightnesses of the objects.  We use the MLCS and dm15 methods
as prescribed by \citet{riess.etal.1998} and
\citet{germany.2001}, respectively.  The MLCS fits the data of \sna\
with a multicolor empirical model that has a single parameter,
luminosity.  This model is matched to the photometry of a \sna,
fitting for the SN luminosity (which is largely a function of the
shape of the light curve, but also its color evolution), time of
maximum light, and reddening.  The empirical model used here was
trained on a large set of well observed \sna\ in the Hubble flow,
using these objects' Hubble velocities as independent measures of
their relative distances and using a quadratic relation between light
curve shape and luminosity \citep{riess.etal.1998}.

An alternate way of measuring \sna\ distances is through the dm15
method
\citep{phillips.1993,hamuy.etal.1996,phillips.etal.1999,germany.2001}.
Here we use the implementation of \citet{germany.2001} because of the
availability of programs, but we would expect similar results if we
were to use the recent work of \citet{phillips.etal.1999}.  Our
implementation of dm15 uses well observed \sna\ (15 objects) to define
a series of multicolor template light curves.  These templates are
labeled by their dm15, the amount that the light curves fall in $m_B$
in the 15 days after maximum light.  Applying each of these templates
to more than 60 objects residing beyond 3000 \kms, we find all
template light curves that give $\chi^2_\nu$ (per degree of freedom)
with a probability $P<95\%$, while fitting for extinction and time of
maximum.  We then use the range of labeled dm15 values and the Hubble
distance to create a diagram as per \citet{phillips.1993}, which plots
dm15 versus absolute magnitude.  In making this diagram, we keep only
those objects that are consistent with no reddening, and find a
similar set of relations for $B$, $V$, $R$, and $I$ as
\citet{phillips.etal.1999}.  To measure the distance to a SN, we then
apply a fitting procedure as above, finding the best fitting SN
template and reddening, and estimating errors by marginalizing over
all SN templates, reddenings, and times of maximum, that provide a fit
with a $\chi^2_\nu$ for $68.3\%$ confidence.

With these two methods, an arbitrary Hubble constant was chosen to
train the methods using the Hubble Flow, and therefore the distances
provided are not absolutely calibrated.  To change to an absolute
scale, we calculate a \dmod\ offset for each method by comparing the
\sna\ distances with their host galaxy Cepheid counterparts.  The
respective offsets are then applied to the SN distances for each
method, providing a set of distances which are tied to the Cepheid
distance scale.

The MLCS and dm15 distances are listed in Table~\ref{tbldmodorig} on the
usual \sna\ zero point calibration.
For MLCS, the \sna\ zero point was calculated from distances based on
the S\&S \citep{parodi.etal.2000} Cepheid distance compilation; however,
for dm15, \citet{germany.2001} used the tabulation of
\citet{suntzeff.etal.1999}.  This results in a 10\% difference in the
Hubble constant---$H_0=65$ (MLCS) and $H_0=70$ (dm15)---but does not
reflect a discrepancy in the methods but, rather, the different
Cepheid distance tabulations.  This issue is treated in much more
detail in \S\ref{secabscal}.  For Table~\ref{tbldmodorig},
we have shifted the dm15 moduli from \citet{germany.2001} to use
the S\&S zero point in order to make comparison between dm15 and MLCS
distances straightforward.  (Once again, we emphasize that the 
SBF and \sna\ distances in Table~\ref{tbldmodorig} are tied to
disparate sets of Cepheid distances and have not yet been homogenized.)

Direct comparison of the dm15 and MLCS distances shows an rms scatter
of 0.23~mag, indicating that the rms scatter of either method
(0.16~mag if taken in equal quadrature) is similar to the rms scatter
measured from the scatter in the Hubble flow and of either method with
SBF.  If we discard SN2000cx which gives a very poor fit by either
method and is nearly a $4\,\sigma$ outlier in the comparison, we find
an rms of 0.18~mag between the methods, implying a dispersion of
0.18~to 0.13~mag by either method depending on their degree of
correlation.  The \sna\ distance methods may benefit from further
refinement, but this will require a larger data set to separate out
true empirical correlations from the underlying noise.

\begin{figure}[t]
\epsscale{0.73}
\plotone{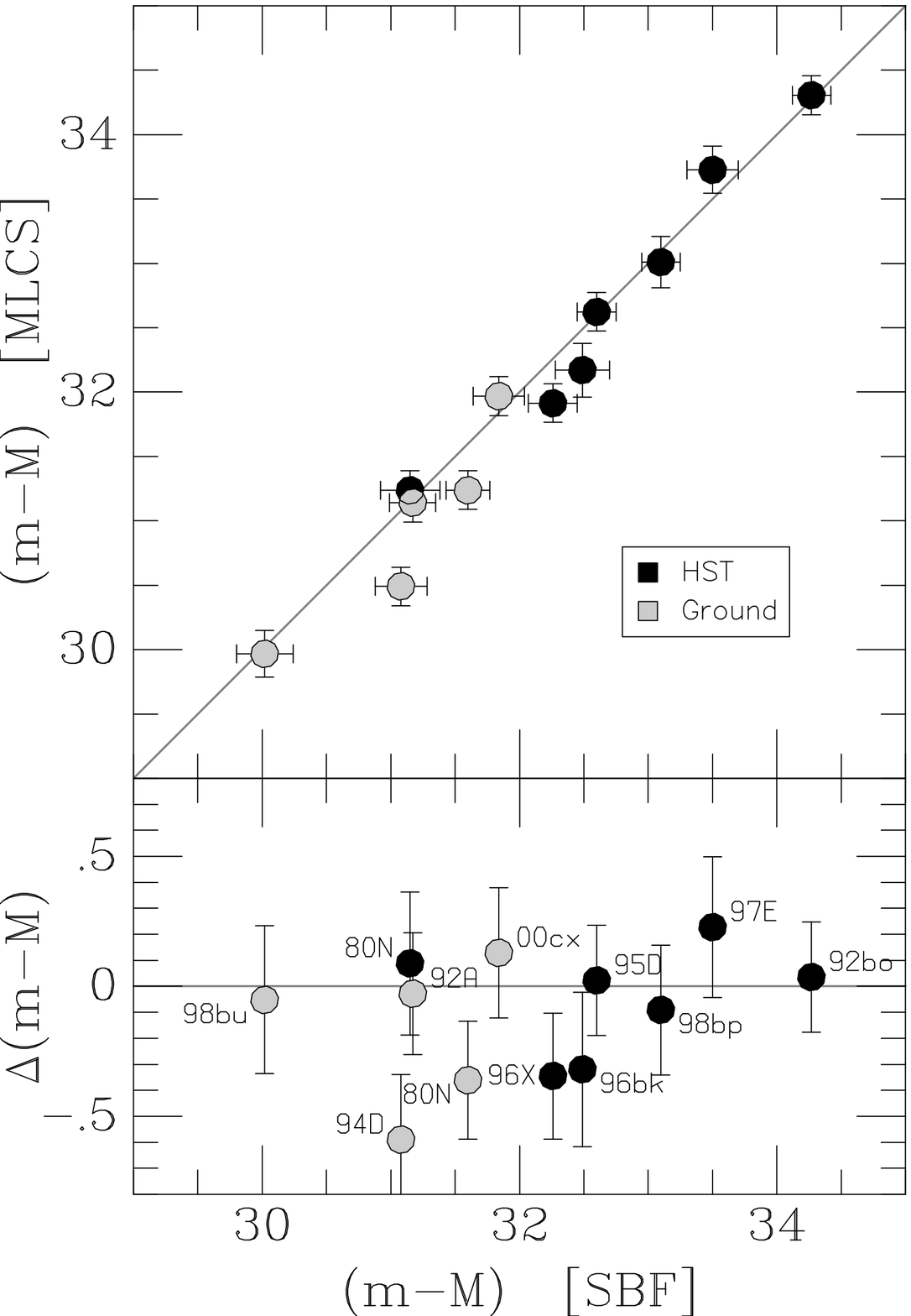}
\caption{\sna\ distance moduli from the MLCS method vs.\ SBF.
The filled circles are \hst\ data, and gray circles are the
ground-based data.  
Both methods have had their zero points calibrated from the Cepheid
distances of \protect{\citet{freedman.etal.2001}}.
\label{comparemlcs}}
\end{figure}

\begin{figure}[t]
\epsscale{0.73}
\plotone{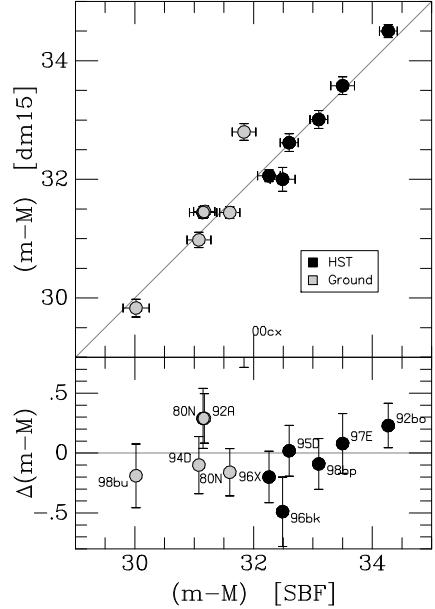}
\caption{\sna\ distance moduli from the dm15 method vs.\ SBF.
The filled circles are \hst\ data, and gray circles are the
ground-based data.
Both methods have had their zero points calibrated from the Cepheid
distances of \protect{\citet{freedman.etal.2001}}.
Note that the HST point for SN80N and the ground-based point
for SN92A are nearly coincident.
Also, the outlier SN00cx is off the top of the residual
plot, although its position is clear from the position of
its label.
\label{comparedm15}}
\end{figure}

\section{SBF Versus \sna\ Distances}
\label{sec:sbfvssna}

In this section we look first at the direct galaxy-to-galaxy
comparison of SBF and \sna\ relative distances and discuss several of
the measurements in detail.  Then, we discuss the absolute calibration
of SBF and \sna\ distances and quantify how the past calibrations have
led to different values of \ho.  Moreover, we address some of the
implications of the different choices that one has in calibrating
these methods.

\subsection{Direct Galaxy-to-Galaxy Comparison}
\label{secreldist}

In order to compare the relative distances from SBF and \sna, it is
convenient to move to a common calibration, either by tying directly
to the Hubble flow and representing the distances in \kms, or by tying
to a homogeneous set of distance calibrators.
Figures~\ref{comparemlcs} and~\ref{comparedm15} show the comparison
between the distance moduli obtained from the SBF and \sna\ methods
as calibrated by the Cepheid distances of
\citet{freedman.etal.2001} for the simplicity of our discussion.  
The distance moduli on this calibration are collected in
Table~\ref{tbldmodkp}.
The \hst\ and ground-based SBF distances are denoted in the figures
by black and gray symbols,  respectively.
It is clear that there is no systematic offset
between the \hst\ and ground distances; and any offset between SBF
and \sna\ distances is quite small (certainly much less than the
offset of $\sim 0.2$~mag that occurs when changing the MLCS zero point
from the Cepheid distances of S\&S to those of
\citealt{freedman.etal.2001}, as discussed in \S~\ref{secabscal}
below).

Although the scatter in these figures is pretty much consistent with
that expected from the error estimates, it is interesting to comment
on outliers.  There is a $1.6\,\sigma$ discrepancy between the
ground-based and \hst\ distances for NGC~1316/SN1980N in Fornax.  The
\sna\ data suggest that the \hst\ distance may be more accurate.
The mean modulus for Fornax in \citetalias{sbf.iv} is 31.49, which is
closer to the \sna\ modulus than either of the two individual
measurements here.

We attempted many times to measure an SBF distance to NGC~2962/SN1995D
from the ground and were never successful, having found power spectra which
did not look sufficiently like that of the PSF.
Examination of the \hst\ data reveals that much of NGC~2962 is covered
by a patchy veil of dust which was not visible at ground-based
resolution.

As noted in~\S\ref{secsbfdist}, we are not completely happy with the
SBF distance and extinction for NGC~5061 because of the inconsistency
between the measured \vi\ color and that inferred from \Nbar.

SN2000cx in NGC~524 is problematic because it does not have a light curve that
follows any template in the dm15 method, nor does it fit the empirical
model of MLCS.  Specifically, the SN rises slowly, and then after a
long time around peak, falls quickly.  The color evolution is also
unusual.  Because this object lies well outside the behavior of the
\sna\ that form the basis of each method, the distances to this object
are highly suspect, and probably should be disregarded.  (See
W.~Li et al., in preparation, for details on SN2000cx.)

SN1996bk in NGC~5308 was discovered approximately one week after maximum, and it
did not have any coverage in the following month which could constrain
the maximum from changes in slope of the light curve.  So, the
possibility of systematic error in the \sna\ distances is
particularly high.

SN1994D in NGC~4526 is an interesting case.  It was observed with
exquisite thoroughness, and is used as a prototypical
``Branch-normal'' \citep*{branch.normal.1993} spectrum.  It had a light
curve which declined rapidly enough that the MLCS fit types it as
rather underluminous.  However, it is also one of the bluest \sna\ ever
observed, particularly in the $U$ band.  This suggests that it may
have been a peculiar supernova with less opacity or more explosion
energy than normal.
\citet{pinto.eastman.2000a,pinto.eastman.2000b} show that this leads
to a light curve which peaks earlier and brighter than a standard
supernova, and then falls more rapidly.  If this is the case, the
normal MLCS analysis would place it at an erroneously small distance,
which could explain the 0.6~mag difference between the MLCS and SBF
distances.  Interestingly, the dm15 method gives an \sna\
distance in good agreement with the SBF distance.

Overall, the relative accuracy of these SBF and \sna\ distances
appears to be outstanding, and the error bars reported by each 
are generally accurate.  The discordant points appear
to be understandable from systematic effects in the
extinction estimates and light-curve template mismatch.

\begin{deluxetable}{llccc}
\tabletypesize{\scriptsize}
\tablewidth{0pt}   
\tablecaption{Distance Moduli Calibrated by
\protect{\citet{freedman.etal.2001}} Cepheid Distances}
\tablehead{
\colhead{Galaxy} &
\colhead{\sna} &
\colhead{$(m{-}M)_{\mathrm{SBF}}$} &
\colhead{$(m{-}M)_{\mathrm{MLCS}}$} &
\colhead{$(m{-}M)_{\mathrm{dm15}}$} }
\startdata%
\multicolumn{5}{c}{WFPC2 F814W Data} \\ \tableline
E352-057&1992bo& $34.27\pm0.15$ & $34.31\pm0.15$ & $34.51\pm0.11$\\
N1316 & 1980N &  $31.15\pm0.23$ & $31.24\pm0.15$ & $31.45\pm0.10$\\
N2258 & 1997E &  $33.50\pm0.15$ & $33.73\pm0.18$ & $33.59\pm0.15$\\
N2962 & 1995D &  $32.60\pm0.15$ & $32.62\pm0.15$ & $32.63\pm0.15$\\
N5061 & 1996X &  $32.26\pm0.19$ & $31.92\pm0.15$ & $32.07\pm0.10$\\
N5308 & 1996bk&  $32.49\pm0.21$ & $32.17\pm0.21$ & $32.01\pm0.20$\\
N6495 & 1998bp&  $33.10\pm0.15$ & $33.01\pm0.20$ & $33.02\pm0.15$\\
\tableline
\multicolumn{5}{c}{Ground-based SBF Survey Data} \\ \tableline
N0524 & 2000cx& $31.84\pm0.20$ & $31.97\pm0.15$\tablenotemark{a} & $32.81\pm0.14$\tablenotemark{a}\\
N1316 & 1980N & $31.60\pm0.17$ & $31.24\pm0.15$ & $31.45\pm0.10$\\
N1380 & 1992A & $31.17\pm0.18$ & $31.14\pm0.15$ & $31.47\pm0.10$\\
N3368 & 1998bu& $30.02\pm0.22$ & $29.97\pm0.18$ & $29.84\pm0.15$\\
N4374 & 1991bg& $31.26\pm0.11$ &    \nodata     &    \nodata    \\
N4526 & 1994D & $31.08\pm0.20$ & $30.49\pm0.15$ & $30.99\pm0.13$\\
N5128 & 1986G & $28.06\pm0.14$ &    \nodata     & $27.17\pm0.20$\\
\enddata
\tablenotetext{a}{This supernova is a very poor fit (W.~Li et al., in
preparation), and these distances are untrustworthy.  See text.}
\label{tbldmodkp}
\end{deluxetable}

\subsection{Absolute Distance Calibration from Cepheids}
\label{secabscal}

Presently, absolute distances from SBF and \sna\ are tied to one
or another set of \hst\ Cepheid distances. 
Although it is possible to calibrate either distance
estimator directly from theory, neither stellar population 
nor white dwarf explosion models have
matured sufficiently to allow us to abandon Cepheid calibration.  (We
briefly discuss the status of theoretical calibrations in the
following section.)

\begin{figure}[t]
\epsscale{0.9}
\plotone{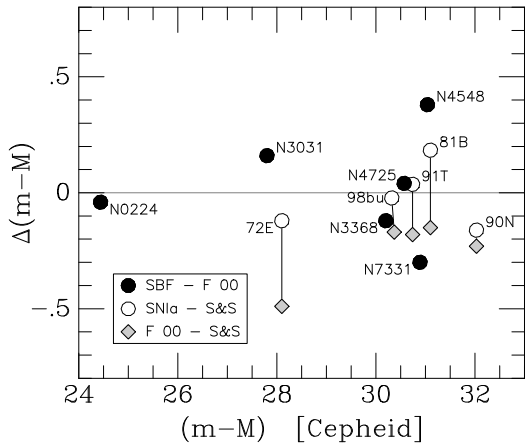}
\caption{Differences in distance moduli to SBF and \sna\ Cepheid
calibrator galaxies vs.\ Cepheid distance moduli.  Black circles are
SBF calibrators on the KP \protect{\citetalias{ferrarese.etal.2000}}
Cepheid scale, i.e., SBF minus
\protect{\citetalias{ferrarese.etal.2000}} modulus vs.\
\protect{\citetalias{ferrarese.etal.2000}} Cepheid distance.  Open
circles are MLCS \sna\ calibrators on S\&S Cepheid scale, i.e., \sna\
minus S\&S modulus vs.\ S\&S Cepheid distance.  Gray diamonds are the
KP \protect{\citetalias{ferrarese.etal.2000}} Cepheids vs.\ the S\&S
Cepheids, i.e., differences in KP
\protect{\citetalias{ferrarese.etal.2000}} and S\&S Cepheids vs.\ the
S\&S Cepheid distances.  The dm15 calibrators show a similar offset
from zero for the \protect{\citetalias{ferrarese.etal.2000}} vs.\ S\&S
calibrators.  SN72E and SN91T have had Cepheid distances measured and
are sometimes used as \sna\ calibrators.  Although we consider both
supernovae to be too untrustworthy to use as calibrators, we plot them
to highlight the \protect{\citetalias{ferrarese.etal.2000}} vs. S\&S
difference.
\label{cepheid}}
\end{figure}

Table~\ref{tblcephcal} lists the Cepheid galaxies chosen to calibrate SBF
and \sna\ and their distance moduli from the collections discussed above.
Figure~\ref{cepheid} illustrates how the differences between SBF and
\sna\ distances originated in the Cepheid calibrations.  The SBF
distances are all tied to the KP Cepheid distances from
\citetalias{ferrarese.etal.2000} according to the \citetalias{sbf.ii}
calibration.  This calibration uses only direct galaxy-to-galaxy
distance matches, from SBF measurements in the bulges of spirals with
Cepheid distances, rather than a potentially troublesome
galaxy-to-group fit.  The SBF calibrators are represented by the black
circles in the diagram.  The \sna\ calibrators, using the S\&S Cepheid
distances, are represented by the open circles in the diagram.
Finally, the gray diamonds show the KP Cepheid distances from
\citetalias{ferrarese.etal.2000} plotted against the S\&S Cepheid
distances; connecting lines emphasize the difference in the two sets
of Cepheid distances.  The result is that the mean offset between the
``long'' (S\&S) and ``short'' (KP) Cepheid distances to these SN
calibrators is the source of past differences between {\ho} values
derived from the SBF and \sna\ methods.  As the recent work by
the KP group (\citetalias{ferrarese.etal.2000};
\citealp{gibson.etal.2000,freedman.etal.2001})
has found consistent values of $H_0$ from SBF and \sna\ when
homogeneously calibrated, a different result would have been surprising.

As listed in Table~\ref{tbldmodorig}, the median difference for 12
galaxies between \sna\ and SBF distance moduli on the original
zero points is 0.23~mag for both the SBF-MLCS and SBF-dm15 comparisons.
The rms scatter in the differences, judged
from the 1/6 and 5/6 points in the distributions, are 0.22~mag and
0.24~mag respectively.
This offset would increase to 0.48 magnitude if we compared to the \sna\
calibration of
\citet{parodi.etal.2000}, but we do not use photographic \sna\ light
curves which we consider to be of questionable value in setting the
\sna\ zero point, nor do we apply some of the corrections
advocated by \citet{parodi.etal.2000}.

If we recalculate our SBF and \sna\ zero points using consistent sets
of Cepheid distances, we find that \citet{freedman.etal.2001} implies
a change of
$-0.06$~mag in SBF moduli with respect to Table~\ref{tbldmodorig} 
(\emph{i.e.,} closer), $-0.33$~mag for MLCS, and $-0.32$~mag for dm15.  
Use of \citetalias{ferrarese.etal.2000} implies offsets of 
$0.00$~mag for SBF,  $-0.25$~mag for MLCS, and $-0.24$~mag for dm15,
and use of \citet{gibson.stetson.2001} causes offsets of
$-0.11$~mag for SBF, $-0.36$~mag for MLCS, and $-0.40$~mag for dm15.
Therefore, using the distances given in \citet{freedman.etal.2001} (as
we have in \S~\ref{secreldist} above and in Table~\ref{tbldmodkp}),
the median difference between \sna\ and SBF distance moduli for the
same galaxies becomes $0.00$~mag for MLCS and $-0.03$~mag for dm15.
Alternatively, using
\citetalias{ferrarese.etal.2000} Cepheid distances yields differences
of $-0.03$~mag for MLCS and $+0.01$~mag for dm15.
In summary, any large disagreement in the zero point for either SBF or
\sna\ distances originates from a disjoint set of Cepheids, each
reduced by a completely separate group.

While this explanation seems simple, we should
not understate the complexity of all the issues involved here.
Although describing the S\&S and KP Cepheid
collections as respectively providing ``long'' and ``short'' scales
was useful for describing our observations, the details
suggest that some of this may have been good fortune.  First, the
choices of Cepheid host galaxies used to calibrate the MLCS and dm15
\sna\ methods will undoubtedly remain in debate among all the
practitioners in the field, and these choices directly affect how well
matched is any particular SBF calibration to a particular \sna\
calibration.  
We note that while we found a 0.23~mag median difference
between S\&S and KP Cepheid distances for
the subset of galaxies chosen, this differs considerably from the mean
offset, and other subsets of common galaxies can produce smaller
offsets of around 0.04~mag, especially when each
galaxy's distance is based only on the same set of Cepheids for both
techniques \citep{parodi.etal.2000}.
In fact, \citet{parodi.etal.2000} have disputed the claim that the KP 
Cepheid distances for the \sna\ calibrators form an internally homogeneous
set with the other KP Cepheid galaxies.
However, detailed discussion of the techniques used by the different groups
for measuring Cepheid distances is beyond the scope of this paper.

Unquestionably, the \sna\ and SBF absolute calibrations
are in need of further refinement, and work is underway to obtain
a better direct Cepheid calibration of SBF with \hst.
In the end, we may find that the best calibrations of SBF and \sna\ absolute
distances will require something different from either of the ones we
have adopted here.

\begin{deluxetable}{llcc}
\tabletypesize{\scriptsize}
\tablewidth{0pt}   
\tablecaption{Distance Moduli of Cepheid Calibrators}
\tablehead{
\colhead{Galaxy} &
\colhead{\sna} &
\colhead{\citetalias{ferrarese.etal.2000}} &
\colhead{\citet{parodi.etal.2000}} }
\startdata%
\multicolumn{4}{c}{SBF Calibrators} \\ \tableline
N0224 &\nodata& $24.44\pm0.10$ & \nodata        \\
N3031 &\nodata& $27.80\pm0.08$ & \nodata        \\
N3368 & 1998bu& $30.20\pm0.10$ & $30.37\pm0.16$ \\
N4548 &\nodata& $31.04\pm0.08$ & \nodata        \\
N4725 &\nodata& $30.57\pm0.08$ & \nodata        \\
N7331 &\nodata& $30.89\pm0.10$ & \nodata        \\
\tableline
\multicolumn{4}{c}{\sna\ Calibrators} \\ \tableline
N3368 & 1998bu& $30.20\pm0.10$ & $30.37\pm0.16$ \\
N4536 & 1981B & $30.95\pm0.07$ & $31.10\pm0.12$ \\
N4639 & 1990N & $31.80\pm0.09$ & $32.03\pm0.22$ \\
\enddata
\label{tblcephcal}
\end{deluxetable}

\subsection{Implications for \ho}

Table~\ref{tblhubble} presents various values of \ho\ which one could
obtain from SBF and \sna\ distances depending on the set of calibrators
chosen.  This table emphasizes that the SBF and \sna\ techniques give
consistent values for \ho\ when the a consistent set of calibrators is
used.

Given that both SBF and \sna\ take their zero points from Cepheids,
which are themselves the main source of disagreement on \ho, 
it might appear that we cannot
contribute much to resolving the \ho\ controversy.  We note, however,
that were we to put SBF on the ``long'' Cepheid scale which has been
used for the \sna\ zero point in the past, the resulting mean SBF
distance modulus for M31/M32 would change from $\mM=24.48\pm0.08$~mag
\citepalias{sbf.iv} to $24.77\pm0.08$~mag.  This conflicts with other
recent Cepheid-independent measures of the M31 distance.  For
instance, \hst\ observations of the horizontal branch (HB) magnitudes
of M31 globular clusters \citep{ajhar.etal.1996,fusipecci.etal.1996},
calibrated from Hipparcos subdwarf parallaxes
\citep[\textit{e.g.,}][]{carretta.etal.2000}, give $\mM\approx24.55$
mag.  Recent results on red clump HB stars
\citep{stanek.garnavich.1998}, globular cluster red giant branch (RGB)
fitting \citep{holland.1998}, and the RGB tip for halo stars
\citep*{durrell.harris.pritchet.2001} all give $\mM=24.47$ mag with
formal errors of $\sim\,$0.1~mag.  The mean magnitude of the M31
globular cluster population \citep*{barmby.huchra.brodie.2001}, when
compared to that of the Milky Way globular clusters $\langle
M_V\rangle=-7.50\pm0.15$ \citep[revised brighter by 0.21 mag according
to the Hipparcos RR~Lyrae zero point]{secker.1992} implies
$\mM=24.34\pm0.19$.  Adopting the brighter
\citet{sandage.tammann.1995} Milky Way $\langle M_V\rangle$
calibration would give $\mM=24.44\pm0.16$.  Thus, combined with these
``Pop~II'' distance estimators, the SBF data for M31/M32 favor the
current SBF calibration from the Key Project Cepheid collections.

In addition, both SBF and \sna\ can be calibrated theoretically to
obtain \ho\ independent of the Cepheids.
\citet{blakeslee.vazdekis.ajhar.2001} and
\citet{liu.charlot.graham.2000} have published SBF calibrations from
stellar population modeling.  The latter calibration is close to the
\citetalias{sbf.ii} ``direct'' (spiral bulge) calibration, while the
former is similar to the group-based SBF calibration.  These model
zero points are in principle tied to the solar model, although they
depend heavily on the Galactic Pop~II distance scale for their stellar
evolution prescriptions.  From \citetalias{sbf.iii},
\citet[transformed to the \citetalias{sbf.ii} ``direct'' zero
point]{lauer.etal.1998}, and \citet{jensen.etal.2001}, we adopt
$\ho=73\pm3$ \kms\ as the current best SBF value, internal errors
only.  Transforming to the \citet{blakeslee.vazdekis.ajhar.2001}
theoretical calibration then gives $\ho=82$, while the
\citet{liu.charlot.graham.2000} calibration gives $\ho=71$; a crude
average ``theoretical \ho\ from SBF'' is then $77\pm7$ \kmsMpc\
($1\,\sigma$ errors).  Clearly, the theoretical SBF calibration favors
the KP Cepheid distance collection over the S\&S one.  The theoretical
calibration for \sna\ \citep{hoeflich.khokhlov.1996} gives
$\ho=67\pm7$ \kmsMpc, which essentially splits the difference between
the two Cepheid collections.

Ultimately, the conflict over the Cepheid distances must be resolved
before a definitive result on \ho\ can be achieved from either SBF or \sna.  
Reassuringly, our results clearly indicate that the two methods will converge
on a \emph{single} definitive \ho\ once this happens.

\begin{deluxetable}{lcccc}
\tablewidth{0pt}   
\tablecaption{Hubble Constants Based on Various Cepheid Collections}
\tablehead{
\colhead{Method} &
\colhead{S\&S} &
\colhead{Gibson \& Stetson} &
\colhead{F00} &
\colhead{Freedman et al.}}
\startdata
SBF  &\nodata&  77   &  73   &  75  \\
MLCS &  65   &  77   &  73   &  76  \\
dm15 &  64   &  77   &  72   &  75
\enddata
\label{tblhubble}
\end{deluxetable}

\section{Conclusions}

We have presented the first direct galaxy-by-galaxy comparison of a
statistically significant sample of SBF and \sna\ distances, where the
latter have been estimated from both the MLCS and dm15 methods.  
The relative agreement between the SBF and \sna\ distances is excellent.
Not only is there \emph{no} evidence for any
scaling error between the two methods, but there \emph{is} evidence that
the small errors predicted by each method are generally accurate, although
with occasional outliers likely due to peculiar \sna\ whose light curves
do not conform to the standard templates.

A systematic offset of 0.23~mag between the absolute SBF and \sna\ distance
moduli has been traced to the discordant sets of Cepheid distances that
have been used for calibrating the methods.  In the past, \sna\
distances were tied to the Cepheid distances from the S\&S collaboration,
while the SBF distances were tied to the KP Cepheid distances.
When a consistent set of Cepheid
calibrating galaxies is used, the distance offset between the
two methods vanishes.  SBF and \sna\ distances therefore
give a common tie to the distant Hubble flow, and if
placed on the same Cepheid distance scale, the two methods
yield the same \ho.  At the same time, we note that
the details of Cepheid distance determination are complex and that our
adopted calibrations of SBF, MLCS, and dm15 may all need to be refined
in the future as the Cepheid observations are better understood.

Our results alone cannot resolve the problem between the ``long''
(S\&S) and ``short'' (KP) Cepheid distances used to calibrate \sna.
However, we have noted that using the ``short'' Cepheid scale allows SBF
distances to M31/M32 to agree with most other distance estimates while
using the ``long'' Cepheid scale would put the SBF distances
uncomfortably at odds with them.  This favors the KP Cepheid scale
(if it is a homogeneous scale), or a value $H_0\gtrsim70$ \kmsMpc.
In addition, theoretical work on SBF
distances also favors the shorter Cepheid scale, while the theoretical
\sna\ scale falls between the long and short Cepheid scales.

It seems clear that any claims of conflicting values for \ho\ from
SBF and \sna, or indeed from any two reliable secondary distance
estimators, most likely result from a failure to reconcile the calibrators
of the two methods rather than from a fundamental problem with either
method.  
The Cepheid distance scale remains uncertain at the $\sim 0.25$~mag level.
Future work must address the problems presented by the Cepheid calibrators
in order to eliminate the remaining disagreement on \ho.  
Using SBF and \sna, we can now place our neighbors M31/M32 with an
accuracy of 5\% in units of the Hubble flow, $57\pm3$~\kms, but until
we know their distance in units of Mpc we will not be able to place
the Hubble flow on a physical scale.

\acknowledgments

Support for this work was provided by NASA through grant number
GO-08212.01-97A from the Space Telescope Science Institute, which is
operated by the Association of Universities for Research in Astronomy,
Inc., under NASA contract NAS5-26555.  We are very grateful to Brian
Barris and Peter Capak for their help with this work, and we thank
Lucas Macri and Abi Saha for valuable comments on the presentation.


\end{document}